\documentclass[journal]{IEEEtran}
\usepackage{amsfonts}
\usepackage{mathtools}
\usepackage{commath}
\usepackage{amssymb}

\usepackage{ifpdf}
\ifpdf
\else
\fi

\ifCLASSINFOpdf
\else
\fi
\usepackage{amsfonts}
\usepackage{mathtools}
\usepackage{commath}
\usepackage{amssymb}
\usepackage[super]{nth}

\usepackage{amsmath}
\usepackage{url}

\begin{document}
%
\title{Statistical Analysis of the LMS Algorithm for Proper and Improper Gaussian Processes}
%
%

\author{Enrique~T.~R.~Pinto~and~Leonardo~S.~Resende}

%
%

\markboth{}%
{}
%



\maketitle

\begin{abstract}
The LMS algorithm is one of the most widely used techniques in adaptive filtering. Accurate modeling of the algorithm in various circumstances is paramount to achieving an efficient adaptive Wiener filter design process. In the recent decades, concerns have been raised on studying improper signals and providing an accurate model of the LMS algorithm for both proper and improper signals. Other models for the LMS algorithm for improper signals available in the scientific literature either make use of the independence assumptions regarding the desired signal and the input signal vector, or are exclusive to proper signals; it is shown that by not considering these assumptions a more general model can be derived. In the presented simulations it is possible to verify that the model introduced in this paper outperforms the other available models.
\end{abstract}


%
\IEEEpeerreviewmaketitle

\section{Introduction}
%
%
%
%
\IEEEPARstart{I}{ntroduced} by Bernard Widrow and Ted Hoff in the 1960s paper \cite{widrow_hoff_lms}, the LMS algorithm is a major landmark in the history of adaptive signal processing algorithms. Its vast domain of applications, ranging from channel equalization to noise reduction, allied to its simplicity of implementation and robust results, has granted it respectable status amongst signal processing techniques even to this day.\par
Even though many variants of the traditional LMS algorithm exist, most of them consistently showing better performance than the original, analysing the fundamental case provides invaluable insight that can be applied on the other cases. \par
Along its history, the complex LMS algorithm, introduced in 1975 by \cite{widrow_complex}, has been modeled in many ways. One of the more acclaimed models was derived by Feuer and Weinstein in \cite{real_lms}, published in 1985 and following up on the initial results from Senne's 1968 Ph.D. dissertation, which sets up the modeling mathematical framework, and on the 1981 paper \cite{horowitz}, which points out the relevance of studying convergence of the weight vector covariance matrix.\par
The previously mentioned models consider the independence assumptions, i.e. the elements of the set $\{\mathbf{x}(n),\, d(n)\}$ are independent from $\{\mathbf{x}(m),\, d(m)\}$ when $n \neq m$. These assumptions are unnecessary for the derivation of the models and increased simplicity one gains from considering them is not substantial. \par 
Furthermore, these models assume the signals to be proper and/or circular. A circular complex random variable have a symmetric probability distribution with respect to arbitrary rotations along the origin in the complex plane. An improper complex random variable is correlated with its complex conjugate. These concepts do not mean the same thing in the general case, however one can quickly develop an intuition on how they are equivalent for Gaussian signals.\par
Improper random processes have only been considered in the modeling of the LMS algorithm recently in the 2009 paper by Mandic and Douglas \cite{mandic_model}. There they draw important results on the basis of a more general model. Their model still considers the independence assumptions, but presents important contributions regarding mean square behavior and its dependence on the noncircularity of the input and desired signals.\par
This paper was developed aiming to further the exploration of the noncircular aspects of the LMS algorithm and to derive a more general model, free of the independence assumptions. One of the relevant achieved results is improved modeling capabilities for large step sizes and small filter lengths.

\section{LMS Algorithm}
\begin{figure}[!htbp]
\centering
\includegraphics[width=8.4cm]{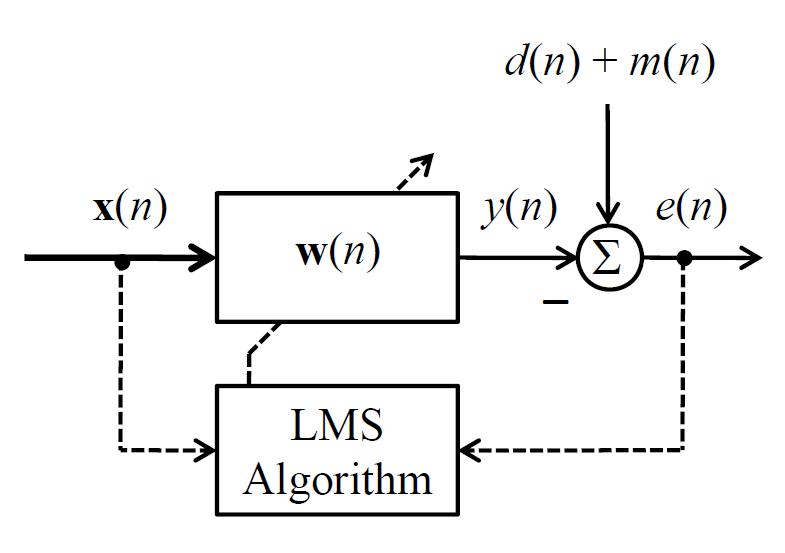}
\caption{Adaptive Wiener filtering scheme represented in block diagram.}
\label{fig:block_filter}
\end{figure}

Figure \ref{fig:block_filter} illustrates Wiener's adaptive filtering scheme in block diagram form \cite{livro_haykin} \cite{livro_widrow}. The LMS algorithm is applied for updating the filter tap-weight vector, according to the recursive equation:
\begin{equation}
    \mathbf{w}(n)=\mathbf{w}(n-1)+\mu e^*(n)\mathbf{x}(n) \label{eqn:w_update}
\end{equation}
where
\begin{equation} \label{eqn:error_signal}
    e(n)=d(n)+m(n)-\mathbf{w^{\mathit{H}}}(n-1)\mathbf{x}(n) 
\end{equation}
is the estimation error at instant $\mathit{n}$, $d(n)$ is the desired signal, $y(n)=\mathbf{w^{\mathit{H}}}(n-1)\mathbf{x}(n)$ is the filter output, $\mathbf{x}(n)$ is the input signal vector, and $\mathit{{m(n)}}$ is a random process added to $d(n)$ for modeling a measurement noise, assumed to be statistically independent of the other aleatory signals. The scalar $\mu$ is the step size parameter. For an $\mathit{N}-1$ order filter, $\mathbf{w}(n)$ and  $\mathbf{x}(n)$ are $\mathit{N}$-element column vectors. Without loss of generality, it is also assumed that all random processes are zero mean.

\subsection*{Optimum Wiener Filtering}
In the mean-square-error (MSE) sense, the vector $\mathbf{w}(n-1)$ is chosen to minimize the following cost function:
\small 
\begin{equation} \label{eqn:MSE}
\begin{split}
    J(\mathbf{w}(n-1))&=E\{|e(n)|^2~|~{\mathbf{w}(n-1)}\}\\
    &=\sigma^2_d+\sigma^2_m-\mathbf{p}^{\mathit{H}}\mathbf{w}(n-1)-\mathbf{w}^{\mathit{H}}(n-1)\mathbf{p}\\
    &~~~+\mathbf{w}^{\mathit{H}}(n-1)\mathbf{R}\mathbf{w}(n-1)
\end{split}
\end{equation}
\normalsize
where $\mathbf{p}=E\{\mathbf{x}(n) d^*(n) \}$ is the cross-correlation vector between $\mathbf{x}(n)$ and $\mathit{{d(n)}}$, $\mathbf{R}=E\{\mathbf{x}(n)\mathbf{x^\mathit{H}}(n) \}$ is the autocorrelation matrix of $\mathbf{x}(n)$, and $\sigma^2_d=E\{|d(n)|^2\}$ and $\sigma^2_m=E\{|m(n)|^2\}$ are the variances of the desired signal and measurement noise, respectively. The optimum solution is
\begin{equation} \label{eqn:wo}
     \mathbf{w}_{\mathit{o}}=\mathbf{R}^{-1}\mathbf{p} 
\end{equation}
Substituting (\ref{eqn:wo}) in (\ref{eqn:MSE}) leads to the minimum MSE (MMSE):
\begin{equation} \label{eqn:MMSE}
    J_{min}=\sigma^2_d+\sigma^2_m-\mathbf{p^{\mathit{H}}}\mathbf{w}_{\mathit{o}}
\end{equation}
Applying (\ref{eqn:wo}) in (\ref{eqn:error_signal}), the expression for the conjugate estimation error at instant $\mathit{n}$ when the filter operates in its optimum condition is given by
\begin{equation} \label{eqn:eoc}
    e_o^*(n)=d^*(n)+m^*(n)-\mathbf{x^{\mathit{H}}}(n)\mathbf{R}^{-1}\mathbf{p} 
\end{equation}
Pre-multiplying both sides of (\ref{eqn:eoc}) by $\mathbf{x}(n)$ and taking the expected value, yields
\begin{equation}
    E\{\mathbf{x}(n)e_o^*(n)\}=\mathbf{0} 
\end{equation}
which permits to verify the orthogonality principle between $\mathbf{x}(n)$ and $\mathit{e_o(n)}$ of the MMSE estimator.
On the other hand, considering that $\mathbf{x}(n)$, $\mathit{{d(n)}}$  and $\mathit{{m(n)}}$ are improper processes, it follows that
\begin{equation} \label{eqn:Exeo}
    E\{\mathbf{x}(n)e_o(n)\}=\mathbf{q}-\mathbf{C}\mathbf{w^*_{\mathit{o}}} 
\end{equation}
where
$\mathbf{q}=E\{\mathbf{x}(n)d(n)\}$ is the pseudo-cross-correlation vector between $\mathbf{x}(n)$ and $\mathit{{d(n)}}$, and $\mathbf{C}=E\{\mathbf{x}(n)\mathbf{x}^T(n)\}$ is the pseudo-autocorrelation matrix of $\mathbf{x}(n)$.
It reveals that the orthogonality principle between $\mathbf{x}(n)$ and $\mathit{e_o^*(n)}$ is not satisfied by the MMSE estimator. In other words, the strictly linear mean-square estimator cannot exploit the full second-order statistics of the signals \cite{picinbono}.

\section{Statistical Analysis}
\subsection{Simplifying Assumptions}
For mathematical tractability, the analysis is performed under the following set of simplifying assumptions:
\begin{description}
\item[A1]$\mathbf{x}(n)$ and $\mathit{{d(n)}}$ are zero-mean complex-valued Gaussian random processes;
\item[A2]$\mathit{{m(n)}}$ is a zero-mean complex-valued white Gaussian random process, which is statistically independent of any other aleatory signal;    
\item[A3]The statistical dependence between $\mathbf{w}(n-1)$ and $\mathbf{x}(n)$ can be neglected.
\end{description}

\subsection{Mean Weight Vector Behavior}
Taking the expectation on both sides of (\ref{eqn:w_update}), and using the above simplifying assumptions, leads to the recursive equation for the transient behavior of $E\{\mathbf{w}(n)\}$:
\begin{equation}
     E\{\mathbf{w}(n)\}=[\mathbf{I}-\mu\mathbf{R}]E\{\mathbf{w}(n-1)\}+ \mu\mathbf{p} \label{eqn:Ew_update}   
\end{equation}
where $\mathbf{I}$ denotes an $\mathit{N} \times \mathit{N}$ identity matrix. \par
The steady state behavior of $E\{\mathbf{w}(n)\}$ can be obtained from (\ref{eqn:Ew_update}), assuming that the LMS algorithm converges when $n\to\infty$:
\begin{equation} \label{eqn:w_infty}
\begin{split}
     \mathbf{w_{\infty}}& \triangleq \lim_{n\to\infty} E\{\mathbf{w}(n)\}   
     \\&=\mathbf{R}^{-1}\mathbf{p} 
\end{split}
\end{equation}
This means that the LMS algorithm converges to the optimum solution in (\ref{eqn:wo}), being the minimum MSE given by (\ref{eqn:MMSE}).

\subsection{Mean Weight Error Vector Behavior}
Defining the $\mathit{N} \times 1$ weight error vector at instant $\mathit{n}$:
\begin{equation}
    \mathbf{v}(n) \triangleq \mathbf{w}(n) - \mathbf{w_{\infty}}
    \label{eqn:Weight_error}   
\end{equation}
from (\ref{eqn:w_update}), yields
\begin{equation}
     \mathbf{v}(n)=\mathbf{v}(n-1)+\mu e^*(n) \mathbf{x}(n) \label{eqn:v_update}   
\end{equation}
Taking the expectation of (\ref{eqn:v_update}), under the simplifying assumptions again, leads to the well known result
\begin{equation}
\begin{split}
    E\{\mathbf{v}(n)\}&=[\mathbf{I}-\mu\mathbf{R}]E\{\mathbf{v}(n-1)\} \\ &=[\mathbf{I}-\mu\mathbf{R}]^{n}E\{\mathbf{v}(0)\} \label{eqn:Ev_update}    
\end{split}
\end{equation}
Hence, convergence in the mean of $\mathbf{w}(n)$ requires all the eigenvalues of $\mathbf{I}-\mu\mathbf{R}$ to be inside the unit circle. In this case, the solution  $\mathbf{w}(n)$ becomes unbiased as $n\to\infty$. Now, using the eigendecomposition $\mathbf{R}=\mathbf{Q \Lambda Q}^H$, where $\mathbf{\Lambda}=diag[\lambda_1,\lambda_2,...,\lambda_N]$ is the diagonal matrix with the eigenvalues of $\mathbf{R}$, and $\mathbf{Q}$ is the unitary matrix consisting of the associated eigenvectors ($\mathbf{QQ^{\mathit{H}}}=\mathbf{I}$), (\ref{eqn:Ev_update}) yields
\begin{equation}
    E\{\mathbf{v}(n)\}=\mathbf{Q}[\mathbf{I}-\mu\mathbf{\Lambda}]^{n}\mathbf{Q}^HE\{\mathbf{v}(0)\} \label{eqn:EigenEv_update}
\end{equation}
From (\ref{eqn:EigenEv_update}), an upper limit for $\mu$ ensuring the convergence of the LMS algorithm can be determined:
\begin{equation}
    0<\mu\ll\frac{1}{\lambda_{max}}   
\end{equation}
where $\lambda_{max}$ is the largest eigenvalue of $\mathbf{R}$, or
\begin{equation} \label{eqn:mi_limit}
    0<\mu<\frac{1}{tr[\mathbf{R}]}
\end{equation}
with $tr[\mathbf{R}]=\sum_{i=1}^N\lambda_i$, and $tr[\bullet]$ denoting the trace of a square matrix $\bullet$.
The time constant of each convergence natural mode is
\begin{equation}
    \tau_i=\frac{1}{ln(1-\mu\lambda_i)}   
\end{equation}
for $\mathit{i}=1, 2,...,N$. Then, for $\mu\lambda_i\ll1$,
\begin{equation} 
    \tau_i\approx\frac{1}{\mu\lambda_i}   
\end{equation}
So, the behavior of the mean weight error vector is dictated only by the eigenvalues of $\mathbf{R}$, regardless of whether $\mathbf{x}(n)$ is proper or improper.

\subsection{Mean Square Error Behavior}
From (\ref{eqn:error_signal}) and (\ref{eqn:Weight_error}), under the simplifying assumptions once again, taking first the conditional expectation, for a given $\mathbf{v}(n-1)$, and after by averaging over $\mathbf{v}(n-1)$, leads to
\begin{equation} \label{eqn:MSE_update}
\begin{split}
    J(\mathit{n})&=E\{|e(n)|^2\}
    \\&=J_{min}+J_{ex}(\mathit{n})
\end{split}
\end{equation}
where $\mathit{J_{min}}$ is the minimum MSE in (\ref{eqn:MMSE}),
\begin{equation} \label{eq:Jex}    J_{ex}(\mathit{n})=tr[\mathbf{R}\mathbf{V}(n-1)]
\end{equation}
accounts for the excess MSE at time $\mathit{n}$ of the stochastic gradient-based adaptation technique, and
\begin{equation} \label{eq:V(n-1)}
\mathbf{V}(n-1)=E\{\mathbf{v}(n-1)\mathbf{v^{\mathit{H}}}(n-1)\}
\end{equation}
is the weight error correlation matrix.

\subsection{Excess MSE}
Equations (\ref{eqn:v_update}) and (\ref{eq:V(n-1)}) yield
\begin{gather}
\begin{aligned}
\mathbf{V}(n-1)=\mathbf{V}(n-2)+\mu[\mathbf{A}(n-1)+ \\ \mathbf{A}^{\mathit{H}}(n-1)]+\mu^{2}\mathbf{B}(n-1)  \label{eq:V_update} 
\end{aligned}\\
\mathbf{A}(n-1)=E\{\mathbf{v}(n-2)\mathbf{x^{\mathit{H}}}(n-1)e(n-1)\}   \label{eq:A(n-1)} \\
\mathbf{B}(n-1)=E\{\mathbf{x}(n-1)\mathbf{x^{\mathit{H}}}(n-1)e(n-1)e^*(n-1)\} \label{eq:B(n-1)}
\end{gather}

From (\ref{eqn:error_signal}) and (\ref{eqn:Weight_error}), based on the assumptions, it follows that
\begin{equation} \label{eq:A_update}
\mathbf{A}(n-1)=-\mathbf{V}(n-2)\mathbf{R}
\end{equation}


Further developing (\ref{eq:B(n-1)}) results in
\begin{gather} 
   \mathbf{B}(n-1)=J(\mathit{n}-1)\mathbf{R}+\mathbf{U}(n-1)+\mathbf{S}(n-1) \label{eq:B_update} \\
   \begin{split}
      \mathbf{U}(n-1)=  E\{\mathbf{x}(n-1)e^*(n-1)\} \cdot \\ \cdot E\{e(n-1)\mathbf{x^{\mathit{H}}}(n-1)\} 
   \end{split}
    \\
    \begin{split}
       \mathbf{S}(n-1)= E\{\mathbf{x}(n-1)e(n-1)\}\cdot \\
        \cdot E\{e^*(n-1)\mathbf{x^{\mathit{H}}}(n-1)\} 
    \end{split}
\end{gather}

Using A1-A3, (\ref{eqn:error_signal}) and (\ref{eqn:Weight_error}), yields
\begin{equation} \label{eq:U(n-1)}
\mathbf{U}(n-1)=\mathbf{R}\mathbf{V}(n-2)\mathbf{R}
\end{equation}
Now, taking also into account that $\mathbf{x}(n)$, $\mathit{{d(n)}}$  and $\mathit{{m(n)}}$ are improper processes, leads to
\begin{equation} \label{eq:S(n-1)}
\begin{split}
\mathbf{S}(n-1)=&\mathbf{k}\mathbf{k^{\mathit{H}}}-\mathbf{k}E\{\mathbf{v}^{\mathit{T}}(n-2)\}\mathbf{C}^{\mathit{H}}\\
&-\mathbf{C}E\{\mathbf{v}^*(n-2)\}\mathbf{k^{\mathit{H}}}+\mathbf{C}\mathbf{V}^*(n-2)\mathbf{C}^{\mathit{H}}
\end{split}
\end{equation}
where
\begin{equation} \label{eq:k}
\mathbf{k}=\mathbf{q}-\mathbf{C}\mathbf{w^*_{\infty}}
\end{equation}
is the pseudo-cross-correlation vector  between $\mathbf{x}(n)$ and $\mathit{e_o(n)}$ in (\ref{eqn:Exeo}).
Equations (\ref{eq:S(n-1)}), (\ref{eq:U(n-1)}), (\ref{eq:B_update}), (\ref{eq:A_update}), (\ref{eq:V_update}) and (\ref{eq:Jex}) summarize the recursive expression for updating $J_{ex}(\mathit{n})$ and, consequently, the MSE behavior in (\ref{eqn:MSE_update}). The model proposed in this paper follows the update equations synthesized below
\begin{gather}
    \mathbf{\overline{v}}(n)=[\mathbf{I}-\mu\mathbf{R}]\mathbf{\overline{v}}(n-1) \\
    \begin{split}
    \mathbf{V}(n)=\mathbf{V}(n-1) - \mu(\mathbf{RV}(n-1) + \mathbf{V}(n-1)\mathbf{R})+ \\
      +\mu^2(J(n)\mathbf{R} + \mathbf{R}\mathbf{V}(n-1)\mathbf{R} + \mathbf{kk}^H - \mathbf{k}\overline{\mathbf{v}}^T(n-1)\mathbf{C^*} - \\
      -\mathbf{C}\overline{\mathbf{v}}^*(n-1)\mathbf{k}^H + \mathbf{C}\mathbf{V}^*(n-1)\mathbf{C^*} )
    \end{split}
    \\
    J(n)=J_{min}+tr\{\mathbf{RK}(n)\}
\end{gather}
For notation cleanness $E\{\mathbf{v}(n-1)\}$ has been rewritten as $\overline{\mathbf{v}}(n-1)$.

\section{Comparison between the proposed model and similar existing models} \label{sec:comp}
Previously presented models of the LMS algorithm's MSE make use of some simplifying assumptions that are not always true, but may under particular circumstances provide a reasonable estimate the actual behavior. \par
For instance, \cite{horowitz} devises a model assuming circular signals; \cite{real_lms} proposes a model for real valued signals (i.e. $\mathbf{R}=\mathbf{C}=\mathbf{C^*}$). The most complete model currently available in the literature is presented in \cite{mandic_model}. However, its derivation considers the pair $\{d(n_1),\mathbf{x}(n_1)\}$ being independent of $\{d(n_2),\mathbf{x}(n_2)\}$ when $n_1 \neq n_2$, also called the \textit{independence assumptions}. Since both the model in \cite{mandic_model} and the model here proposed consider $d(n)$ and $\mathbf{x}(n)$ to be jointly Gaussian random variables in order to allow the use of Gaussian fourth moment factoring theorem, it is reasonable to state that the model presented in this paper is more general. \par
The remainder of the analysis in this section will be made in comparison to model in \cite{mandic_model}, due to it being the one that has the most theoretical proximity to the proposed model and has the best results amongst the other models in approximating the actual MSE behaviour.\par
Differences between the two models may occur only on the $\mu^2$ terms in the $\mathbf{K}(n)$ matrix update equations, more precisely those containing $\mathbf{k}$. In fact, if $\mathbf{k}=\mathbf{0}$, then the expressions of the proposed model reduce to those of \cite{mandic_model}. Therefore to properly visualize the differences between models a large step size is required, otherwise the $\mu^2$ term's influence is negligible. \par
It is illuminating to analyse two special cases for the matrices $\mathbf{R}$ and $\mathbf{C}$. Restating the model update equations

\begin{gather}
    \mathbf{\overline{v}}(n)=[\mathbf{I}-\mu\mathbf{R}]\mathbf{\overline{v}}(n-1) \\
    \begin{split}
    \mathbf{K}(n)=\mathbf{K}(n-1) - \mu(\mathbf{RV}(n-1) + \mathbf{V}(n-1)\mathbf{R})+ \\
      +\mu^2(J(n)\mathbf{R} + \mathbf{R}\mathbf{V}(n-1)\mathbf{R} + \mathbf{kk}^H - \mathbf{k}\overline{\mathbf{v}}^T(n-1)\mathbf{C^*} - \\
      -\mathbf{C}\overline{\mathbf{v}}^*(n-1)\mathbf{k}^H + \mathbf{C}\mathbf{V}^*(n-1)\mathbf{C^*} )
    \end{split}
    \\
    J(n)=J_{min}+tr\{\mathbf{RV}(n)\}
\end{gather}
\normalsize
\subsection{Uncorrelated Non-Circular Input Signal}
The first case to be analysed has covariance and pseudocovariance matrices of the form
\begin{gather}
        \mathbf{R}=\sigma^2 \mathbf{I}\\
        \mathbf{Q}^H\mathbf{C}\mathbf{Q^*}=\sigma^2 \mathbf{\Lambda}
\end{gather}
Where $\mathbf{\Lambda}$ is a diagonal matrix of circularity coefficients.\par
Define $\Tilde{\mathbf{K}}(n)=\mathbf{Q}^H\mathbf{K}(n)\mathbf{Q}$, and define $\Tilde{k}_i$ as the `i'th element of $\mathbf{Q}^H\mathbf{k}$. Taking the limit as $n\xrightarrow{}\infty$ yields the steady state excess MSE expression
\begin{equation}
    J^{ex}_{\infty}=\frac{\mu\sum^N_{i=1}\frac{J_{min}\sigma^2+|\Tilde{k_i}|^2}{2-\mu\sigma^2(1+\lambda^2_i)}}{1-\sum^{N}_{j=1}\frac{\mu\sigma^2}{2-\mu\sigma^2(1+\lambda^2_k)}} \label{eq:j_ex_caso_a}
\end{equation}
Which is remarkably similar to the equivalent expression in \cite{mandic_model}. The difference once again arises in the terms concerning $\mathbf{k}$.\par
Requiring that
\begin{equation}
    1-\sum^{N}_{j=1}\frac{\mu\sigma^2}{2-\mu\sigma^2(1+\lambda^2_k)}>0 \label{eq:condicao_caso_a}
\end{equation}
so that (\ref{eq:j_ex_caso_a}) remains finite returns a new upper bound for the step size. Expanding (\ref{eq:condicao_caso_a}) in order to isolate $\mu$ gives messy expressions that give no immediate contribution to intuition. A simplified form can be obtained by realising that
\begin{equation}
    \sum^{N}_{j=1}\frac{\mu\sigma^2}{2-\mu\sigma^2(1+\lambda^2_k)}<\frac{N\mu\sigma^2}{2-\mu\sigma^2(1+\lambda^2_{max})}
\end{equation}
It can then be shown that
\begin{equation}
    \mu<\frac{2}{\sigma^2(N+1+\lambda^2_{max})} \label{eq:bound_caso_a}
\end{equation}
The upper bound in \ref{eq:bound_caso_a} formalizes a very perceivable aspect of convergence of the LMS algorithm under different circularity conditions. In general, as the input signal becomes more improper the signal, the upper bound for step size decreases. This means that, even if two signals have the same $\mathbf{R}$ and $\mathbf{p}$, but have different circularity coefficients, a step size that converges for one signal may not converge for another with higher circularity coefficients.\par
It is evident that if the approximate expression for the upper bound is not satisfactory, one can always turn to numerical evaluation for a more precise value, as it is a simple optimization problem. 
\subsection{Uniform Non-Circularly Correlated Data}
Now taking 
\begin{gather}
        \mathbf{R}=\mathbf{Q}\mathbf{Q}^H\\
        \mathbf{C}=\lambda \mathbf{Q}\mathbf{Q}^T
\end{gather}
\cite{mandic_model} states that $\mathbf{R}$ and $\mathbf{C}$ may be written on the following form based on the SVD of the Strong Uncorrelating Transform $\mathbf{Q}$
\begin{gather}
    \mathbf{R}=\mathbf{U}\mathbf{\Sigma}^2\mathbf{U}^H\\
        \mathbf{C}=\lambda \mathbf{U}\mathbf{\Sigma}^2\mathbf{U}^T
\end{gather}
where $\mathbf{U}$ is the matrix containing the left singular vectors of $\mathbf{Q}$ and $\mathbf{\Sigma}$ is the diagonal matrix of associated singular values.\par
Similar to Case A, define $\Tilde{\mathbf{K}}(n)=\mathbf{U}^H\mathbf{K}(n)\mathbf{U}$, and define $\Tilde{k}_i$ as the `i'th element of $\mathbf{U}^H\mathbf{k}$. The steady state excess MSE expression resembles (\ref{eq:j_ex_caso_a}) and is then given by
\begin{equation}
    J^{ex}_{\infty}=\frac{\mu\sum^N_{i=1}\frac{J_{min}\sigma^2_i+|\Tilde{k_i}|^2}{2-\mu\sigma^2_i(1+\lambda^2)}}{1-\sum^{N}_{j=1}\frac{\mu\sigma^2_k}{2-\mu\sigma^2_k(1+\lambda^2)}} \label{eq:j_ex_caso_b}
\end{equation}
Looking for an upper bound in much the same way as in the previous case results in the approximate expression
\begin{equation}
    \mu<\frac{2}{\sigma^2_{max}(N+1+\lambda^2)} \label{eq:bound_caso_b}
\end{equation}
Reinforcing the intuition that larger $\lambda$ implies a smaller upper bound for the step size.

\section{Simulations}
To show the effectiveness of the model proposed in this paper, some numerical evaluations of system identification and channel equalization Wiener filtering are presented.\par In both setups non circular white noise is used either as input for both the filter and system to be identified in the case of system identification, or as the original transmitted signal in the channel equalisation setup. Once again the main comparison is drawn relating to \cite{mandic_model}, henceforth also referred to as ``Independence Model". For aforementioned reasons the step sizes in the presented simulations are taken to be as high as possible while the Monte Carlo simulation is still consistently converging.\par
Defining $x[n]=u+jv$, with $u,v \in \mathrm{R}$, and arranging $u$ and $v$ in a vector $\mathbf{z}=\begin{bsmallmatrix} u & v \end{bsmallmatrix}^T$, the PDF of $u$ and $v$ is given by
\small
\begin{equation}
    p_{uv}(u,v)=\frac{1}{2 \pi \, \textrm{det}^{1/2}\mathbf{R}_{zz}} \, \textrm{exp} \left( -\frac{1}{2} \begin{bmatrix} u & v \end{bmatrix} \mathbf{R}^{-1}_{zz} \begin{bmatrix} u \\ v \end{bmatrix}  \right)
\end{equation}
\normalsize
in which $\mathbf{R}_{zz}$ is the autocorrelation matrix of $\mathbf{z}$. $\mathbf{R}_{zz}$ can be expressed as a function of the variances of $u$ and $v$ and their correlation coefficient $\rho_{uv}$.
\begin{equation}
    \mathbf{R}_{zz}=\begin{bmatrix} r_{uu} & \sqrt{r_{uu}} \sqrt{r_{vv}}\rho_{uv} \\
    \sqrt{r_{uu}} \sqrt{r_{vv}}\rho_{uv} & r_{vv}\end{bmatrix}
\end{equation} \par
It is necessary to determine the covariance and pseudocovariance of $x$, as they are used extensively in the models, their expressions are
\begin{gather}
    r_x=r_{uu}+r_{vv}   \\
    q_x=r_{uu}-r_{vv}+j2\sqrt{r_{uu}} \sqrt{r_{vv}}\rho_{uv}
\end{gather}
The \textit{impropriety coefficient} denoted by $\rho_x$, also known as the \textit{complex correlation coefficient}, is the ratio between the pseudocovariance and covariance of $x$. For convenience, the variances of the real and imaginary parts of $x[n]$ are $r_{uu}=r_{vv}=0.1$ in all simulations, so that $\rho_x=j\rho_{uv}$. \par

The input autocorrelation matrix $\mathbf{R}$ and the pseudocorrelation matrix $\mathbf{C}$ in this case are
\begin{gather}
    \mathbf{R}=\sigma^2_x \mathbf{I}  \label{eq:R_ident}\\
    \mathbf{C}=q_x \mathbf{I} \label{eq:C_ident}
\end{gather}
with $q_x=E\{ x(n)x(n) \}$.

\subsection{System Identification}
If the system to be identified were a simple FIR filter with coefficients $\mathbf{f}^H$, under the current setup of non-circular white Gaussian noise input, then $\mathbf{k}=\mathbf{0}$, because $
    \mathbf{q}=\mathbf{C}\mathbf{f}^*$, 
    $\mathbf{C}=q_x\mathbf{I}$, and  
    $\mathbf{w_o}=\mathbf{f}$.
Therefore, this kind of system is of no use in extracting different behavior from the proposed model and the model in \cite{mandic_model}.\par
To circumvent this problem, the system to be identified is a Widely Linear FIR filter of the form
\begin{equation}
    d(n)=\mathbf{f}^H\mathbf{x}(n)+\mathbf{g}^H\mathbf{x}^*(n)
\end{equation}
The system input $x[n]$ here is, in general, non-circular complex Gaussian white noise. The desired output $d(n)$ is corrupted by circular white Gaussian noise $\nu(n)$ with variance $\sigma^2_\nu$\par 
The crosscovariance and pseudo-crosscovariance vectors between input and desired signal, $\mathbf{p}$ and $\mathbf{q}$ respectively, can be obtained by the following relations
\begin{gather}
    \mathbf{p}=\mathbf{Rf} + \mathbf{Cg}     \\ 
    \mathbf{q}=\mathbf{Cf}^* + \mathbf{Rg}^*
\end{gather}\par
Evaluating $\mathbf{k}$ returns
\begin{equation}
    \mathbf{k}=(\mathbf{R}-\mathbf{C}\mathbf{R}^{-*}\mathbf{C}^*)\mathbf{g}^* \label{eq:k_ident}
\end{equation}
which is only zero when the Schur complement of $\mathbf{R}^*$ in the augmented covariance matrix as defined in \cite{schreier_scharf_2010} is zero. It generalizes the case when $\mathbf{g}=0$ (i.e. strictly linear FIR filter) and shows once again that both models are equal if the system is a strictly linear FIR filter. Also, by (\ref{eq:R_ident}), (\ref{eq:C_ident}), and (\ref{eq:k_ident}) it follows that
\begin{equation}
    \mathbf{k}=\sigma^2_x(1-|\rho_x|^2)\mathbf{g}^* \label{eq:k_rho_ident}
\end{equation}
This means that, in the scope of system identification in the form here presented, $\norm{\mathbf{k}}$ becomes progressively smaller as $|\rho|$ becomes larger, causing both models to become progressively closer and achieving equality when $|\rho_x|=1$. \par
\begin{figure}[!htbp]
\centering
\includegraphics[width=8.4cm]{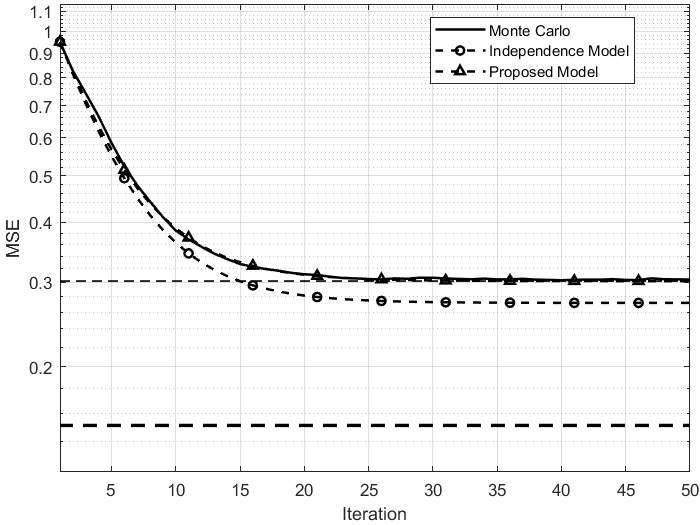}
\caption{System identification simulation for $\mathbf{f_1}$ and $\mathbf{g_1}$ with $\mu=1$, $\sigma^2_\nu=10^{-3}$, $|\rho_x|=0$, and $3\cdot10^5$ simulation runs.}
\label{fig:ident_s1}
\end{figure}

All the simulations of this subsection use a Wiener filter of length $N=4$. Particularly, in the simulation depicted in Fig. \ref{fig:ident_s1}, the widely linear system given by $\mathbf{f_1}=\begin{bsmallmatrix} 1 & j & 1 & j \end{bsmallmatrix}$ and $\mathbf{g_1}=\begin{bsmallmatrix} 0.5j & 0.5 & 0 & 0.5 \end{bsmallmatrix}$ is identified via an input with circular characteristics ($|\rho_x|=0$) and large step size ($\mu=1$). This favors the visualization of the differences between the proposed model and the model that considers the independence assumptions because $\mathbf{k}$ is not negligible. In fact, $\norm{\mathbf{k}}^2=\mathbf{k}^H\mathbf{k}$ may be used as a numerical value to quantify what is intuitively considered to be the ``difference" between the two models. In Fig. \ref{fig:ident_s1}, $\mathbf{k}^H\mathbf{k}=0.030$.\par
The thicker dotted line represents $J_{min}$, the minimum MSE according to the Wiener-Hopf solution (\ref{eqn:MMSE}), and the thinner dotted line is the steady state solution of the Proposed Model adding $J_{min}$ to the expression of the excess error in (\ref{eq:j_ex_caso_a}). Both lines are presented in the simulations to visualize the excess MSE and the accuracy of the expressions.

This simulation gives evidence that the proposed model is more accurate than the Independence Model. The closer proximity of the proposed model and the experimental curve, when compared to the Independence Model, persists in simulations concerning other systems even if it is not as apparent as in Fig. \ref{fig:ident_s1}. The transient responses and the steady state MSE of the Proposed Model are generally closer to the Monte Carlo simulations in relation with the Independence Model.\par
Looking into the numerical values of steady state results confirms what is already noticeable by looking at the graphs. Assuming that in Fig. \ref{fig:ident_s1} all of the curves have already converged sufficiently close to their steady state value by the \nth{100} iteration, the \nth{100} MSE sample average in the $3\cdot 10^6$ Monte Carlo simulation runs of the LMS algorithm is 0.3041, while in the Proposed Model's predicted value is 0.3018, yielding -0.756\% relative error. The Independence Model produces a predicted MSE value of 0.2718, corresponding to a -10.621\% relative error. In this case, the benefit offered by using the newly derived model is clearly significant.\par
\begin{figure}[!htbp]
\centering
\includegraphics[width=8.4cm]{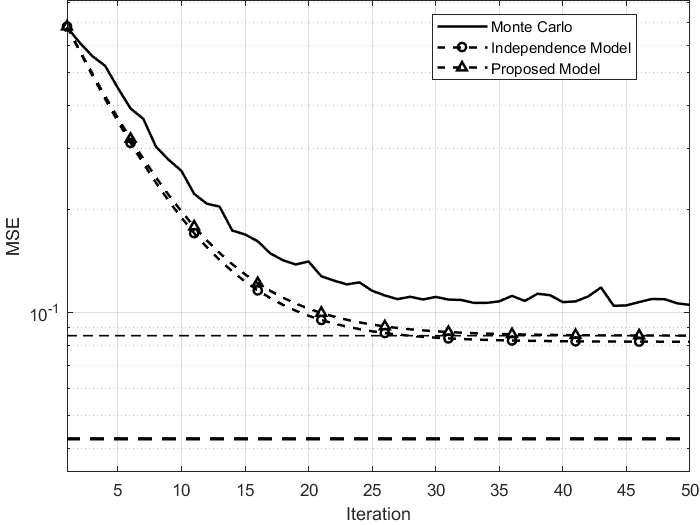}
\caption{System identification simulation for $\mathbf{f_2}$ and $\mathbf{g_2}$ with $\mu=1$, $\sigma^2_\nu=10^{-3}$, $|\rho_x|=0.8$, and $10^5$ simulation runs.}
\label{fig:ident_s2}
\end{figure}

To further illustrate this point and to explore different contexts of the LMS algorithm the simulation in in Fig. \ref{fig:ident_s2} is given. A new system is used to further explore the behavior of the models, in this case $\mathbf{f_2}=\begin{bsmallmatrix} 1 & 0.5j & 0.5 & -1 \end{bsmallmatrix}$ and $\mathbf{g_2}=\begin{bsmallmatrix} 0.2 & 0.5j & 0.5 & -0.2j \end{bsmallmatrix}$ As this simulation explores a more improper input signal ($|\rho_x|=0.8$) it is expected that both models become closer to one another, in fact $\mathbf{k}^H \mathbf{k}=0.003007$, which is about an order of magnitude smaller than the value for the first simulation shown, however the Proposed Model still maintains its better performance. Assuming convergence to steady state by the \nth{50} iteration, The Monte Carlo MSE of $10^5$ runs is 0.1051, while the Proposed Model's prediction is 0.08544, corresponding to a relative error of -18.705\%. The Independence Model's prediction is 0.08204, with relative error of -21.941\%; while both models are not as different as in the previous case, the Proposed Model still shows a more accurate result.

\subsection{Channel Equalization}
Exploring the application of Wiener filtering to channel equalization concerning the transmission of generally non-circular data yields more evidence of the better predictive power of the model introduced in this article. \par
The equalized channel in the simulations is modeled by an FIR filter $\mathbf{f}$ with $M$ complex coefficients. In particular, the channel used for this section is given by $\mathbf{f}^H=\begin{bsmallmatrix}0.3 & -0.5 & -0.7j & 1\end{bsmallmatrix}$, so that the received signal is $\mathbf{x}=\mathbf{w}^H\mathbf{u}(n)+\nu(n)$, in which $\mathbf{u}(n)$ is the vector of the last $M$ transmitted symbols $u(n)$, and $\nu(n)$ is circular AWGN. The desired signal $d(n)$ is the transmitted symbols signal delayed by $\alpha$ samples so as to compensate the equalizer filter length, i.e. $d(n)=u(n-\alpha)$. In the simulations, $u(n)$ is a non-circular white Gaussian signal (as described in the beginning of this section), $\alpha=4$ and the equalizer filter length is $N=5$. \par
In general, the form of matrices $\mathbf{R}$ and $\mathbf{C}$ in channel equalization do not match those discussed in Section \ref{sec:comp}, as a consequence of this, no \textit{closed form} expressions for the steady state MSE are available. As a way to present steady state values for both models these values are approximated by taking the last calculated value for each model.\par

\begin{figure}[!htbp]
\centering
\includegraphics[width=8.4cm]{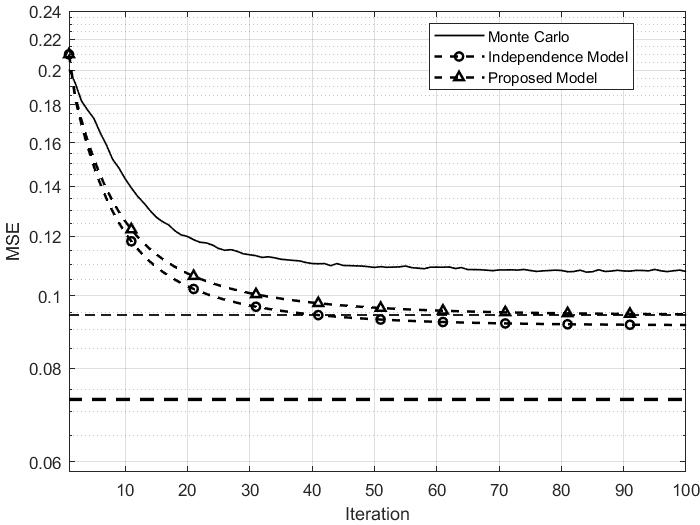}
\caption{Channel equalization simulation for $\mathbf{f}$ with $\mu=0.2$, $\sigma^2_\nu=10^{2}$, $|\rho_x|=1$, and $10^5$ simulation runs.}
\label{fig:equal_1}
\end{figure}

Fig. \ref{fig:equal_1} presents the MSE curves of the LMS algorithm subjected to high impropriety input signal ($|\rho_x|=1$). It is important to realize that the expression in (\ref{eq:k_rho_ident}) does not present valid information regarding channel equalization, that is, both models do not become increasingly close as $|\rho_x|$ increases in the context of channel equalization. As a reference, for channel $\mathbf{f}$, if $|\rho_x|=1$ then $\mathbf{k}^H\mathbf{k}=0.022273$, if $|\rho_x|=0$ then $\mathbf{k}^H\mathbf{k}=0$; which corresponds to the exact opposite general behavior.\par
Analysing the MSE curves in Fig. \ref{fig:equal_1} shows that the Proposed Model is closer to the Monte Carlo curves than the Independence Model during the whole transient. The estimated steady state value of the Proposed Model, despite lack of availability of theoretical expressions, is also reasonably closer to the Monte Carlo value than the Independence Model approximate steady state value. Considering that all curves are sufficiently close to steady state by the \nth{100} sample, the ensemble average of $10^5$ simulation runs returns steady state MSE of 0.1077; the Proposed Model's prediction is 0.09419, corresponding to relative error of -12.544\%; the Independence Model's prediction is 0.09134, with relative error value of -15.190\%. Once again the relative error associated with the Proposed Model is smaller than the relative error of the Independence Model.

\section{Conclusion}
This paper introduced a model for the complex LMS algorithm that generalizes those published in \cite{horowitz}, \cite{real_lms} and \cite{mandic_model}. Although steady state expressions could not be immediately derived for all cases, as a consequence of the autocovariance and pseudocovariance matrices generally not sharing an eigenspace, expressions were provided for some cases of interest. The accuracy of the new model was tested by comparing the simulation results between the proposed model and the model introduced in \cite{mandic_model}. Both models differ only in second order terms, therefore the differences are most perceptible for large step sizes and small filter lengths. The model introduced in this paper is consistently closer to the real LMS Monte Carlo behavior than the other mentioned models.


%




\ifCLASSOPTIONcaptionsoff
  \newpage
\fi



%

\bibliographystyle{unsrt} 
\bibliography{references.bib} 


%








\end{document}